\documentclass[twocolumn,showpacs,preprintnumbers,footinbib]{revtex4-1}
\usepackage{bm,enumerate,dcolumn,tikz,graphicx,color,amsmath,amssymb}
\usepackage{epstopdf}
\usepackage{empheq,xfrac}
\usepackage{capt-of}
\usepackage{pgfplots}

\usepgfplotslibrary{colormaps,external}
\usetikzlibrary{decorations.markings}
\usetikzlibrary{arrows}

\newcommand{\comment}[1]{}
\newcommand{\bra}{\langle}
\newcommand{\ket}{\rangle}

\newcommand{\Rmnum}[1]{\expandafter\@ slowromancap\romannumeral #1@}
\newcommand{\RNum}[1]{\uppercase\expandafter{\romannumeral #1\relax}}
\definecolor{lightblue}{rgb}{.8, .8, 1}
\definecolor{myyellow}{RGB}{254,241,24}
\definecolor{myorange}{RGB}{234,125,1}
 
\usepackage{array}
\newcolumntype{C}[1]{>{\centering\let\newline\\\arraybackslash\hspace{0pt}}m{#1}}
\newcolumntype{L}[1]{>{\raggedright\let\newline\\\arraybackslash\hspace{0pt}}m{#1}}
\newcolumntype{R}[1]{>{\raggedleft\let\newline\\\arraybackslash\hspace{0pt}}m{#1}}

\begin{document}

\title{Possibility of triple magic trapping of clock and Rydberg states of divalent atoms in optical lattices} 
\author{T. Topcu and A. Derevianko}
\affiliation{
$^{1}$Department of Physics, University of Nevada, Reno, Nevada 89557, USA \\
$^{2}$ITAMP, Harvard-Smithsonian Center for Astrophysics, Cambridge, Massachusetts 02138, USA 
}
\date{\today}

\begin{abstract}
We predict the possibility of ``triply-magic'' optical lattice trapping of neutral divalent atoms. In such a lattice, the ${^1}\!S_{0}$ and ${^3}\!P_{0}$ clock states and an additional Rydberg state experience identical optical potentials, fully mitigating detrimental effects of the motional decoherence. In particular, we show that this triply magic trapping condition can be satisfied for Yb atom at optical wavelengths and for various other divalent systems (Ca, Mg, Hg and Sr) in the UV region. We assess the quality of triple magic trapping conditions by estimating the probability of excitation out of the motional ground state as a result of the excitations between the clock and the Rydberg states. We also calculate trapping laser-induced photoionization rates of divalent Rydberg atoms at  magic frequencies. We find that such rates are below the radiative spontaneous-emission rates, due to the presence of Cooper minima in photoionization cross-sections. 
\end{abstract}

% pacs, the Physics and Astronomy Classification Scheme
\pacs{32.80.Ee, 34.20.Cf, 37.10.Jk} 

\maketitle

% introduction 
\section{Introduction}\label{sec:intro}
Recent advances in precision time keeping, quantum information processing (QIP) and many-body simulation rely on trapping cold atoms in optical lattices. In particular, harnessing the rapid on-demand access to strong long-range interactions between trapped Rydberg atoms holds an intriguing promise for realizing scalable quantum computing~\cite{UrbJohHen09,GaeMirWil09,SafWalMol10} and creating massive entanglement~\cite{SafMol09}. Current experimental efforts focus on alkali-metal atoms and encode qubits into the ground state hyperfine manifold. However, clocks using hyperfine structure rely on microwave transitions and they have been far surpassed by clocks that rely on optical transitions (see, {\it e.g.}, micromagic clocks in~\cite{DerKat11}). Here we explore some aspects of a different scheme that utilizes divalent atoms, where we encode the qubit states into the long-lived and well-protected clock states: the ground $^1\!S_0$ and the lowest-energy $^3\!P_0$ state. 

Working with divalent atoms in QIP provide several advantages over alkali metal atoms. For example, Rydberg series perturbation can be exploited to engineer van der Waals interaction strengths between Ry states in a way that deviates from its monotonic $n$-dependence in alkali metal atoms~\cite{TopDer_SeriesPert15}. Furthermore, the trapping potential seen by divalent atoms in an optical lattice results from Ry electron polarizability, which sits on a background of ion core polarizability~\cite{TopDer14}. This leads to ``universal" magic wavelengths (typically below $\sim$1000 nm), which do not depend on the principal quantum number of the Ry state - something that does not exist for alkali metals~\cite{TopDer13}. Recently, a quantum protocol using Yb to realize an entangled network of atomic clocks has been proposed~\cite{KomTopKes16}, owing to its available telecom wavelength transitions. 

In all the QIP applications,  a multitude of decoherence mechanisms must be overcome~\cite{SafWal05}. One of such challenges is the motional  decoherence: if two atomic states $e$ and $g$ experience different optical potentials, there is a lattice laser intensity-dependent differential AC Stark shift between the two levels and also there is motional heating during the population transfer between two internal states. The difference in optical potentials and thereby the motional decoherence can be fully eliminated in magic latices~\cite{KatTakPal03,TakHonHig05}, which are operated at specially chosen ``magic'' lattice laser wavelengths (or frequencies $\omega_m$). Since the optical potentials are proportional to dynamic polarizability, $\alpha(\omega_m)$ the ``magic'' condition reads $\alpha_e(\omega_m) = \alpha_g(\omega_m)$.
 
\begin{figure}[h!tb]
	\begin{center}
				\resizebox{0.9\columnwidth}{!}{\includegraphics[angle=0]{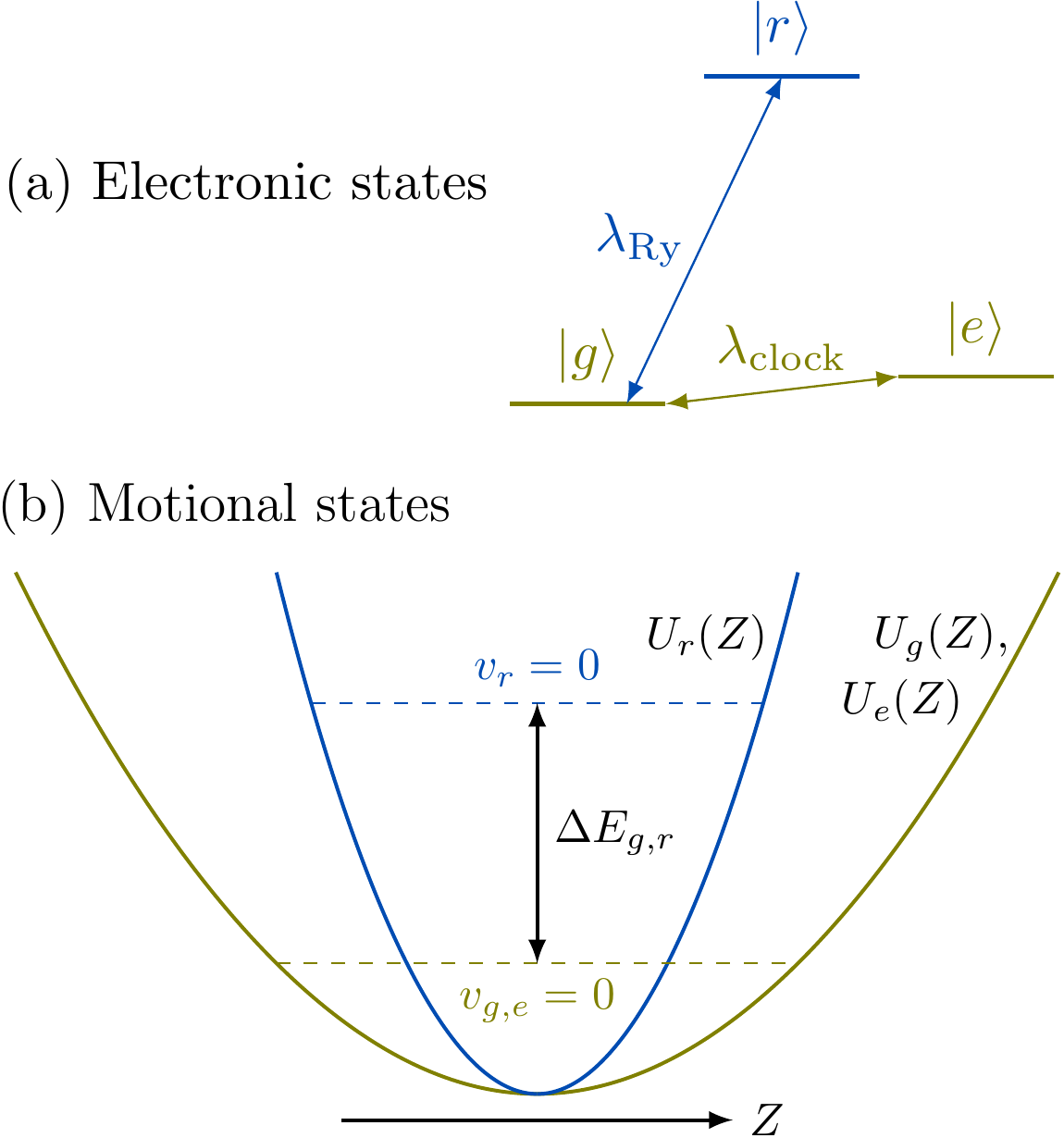}} 
	\end{center}
	\caption{(Color online) (a) Electronic and (b) motional energy levels relevant to excitations between clock and Ry states of an atom trapped in an optical lattice. In general, the clock ($|g\ket$ and $|e\ket$) and Ry ($|r\ket$) atoms experience different optical potentials, $U_{g,e}(Z)$ and $U_{r}(Z)$. %This  configuration is elemental to experiments simulating quantum logic gates using the Ry blockade mechanism to mediate conditional logic. 
	}
	\label{fig:diagram}
\end{figure}

Namely the magic wavelength lattice idea is the key  behind the overwhelming success of optical lattice clocks~\cite{DerKat11}. In this case, the lattice wavelength is chosen so that the ${^1}\!S_{0}$ and ${^3}\!P_{0}$ clock states experience the same optical trapping potential. Here we expand upon this idea and demonstrate that magic trapping conditions can be attained simultaneously for the two clock levels and an additional Ry level (Fig.~\ref{fig:diagram}~(a)). We term this case where all three states (the clock and Ry states), experience identical optical potentials as the ``triple" magic trapping.   Notice that this phenomenon is unique to divalent (and possibly other multi-valent) atoms. 

Earlier attempts~\cite{MorDer12} failed to find ``triple" magic wavelengths for the alkali metal atoms. We demonstrate that such magic wavelengths do exist for divalent atoms.  One of the reasons  is the difference in relative size of the ion  polarizability between multivalent and mono-valent Rydberg atoms. (Ion is the residual singly-charged ion, e.g. for Sr, it is the Sr$^+$ ion.). In alkali (mono-valent) atoms, the core electrons are tightly bound with excitation wavelengths in the UV, while in multi-valent atoms the core is ``softer'' and highly-polarizable with much-longer excitation wavelengths.  As both the core and Ry electron polarizabilities  contribute to the total atomic polarizability,  the core polarizability may play the dominant role in optical trapping of multi-valent atoms.

Another relevant point worth emphasizing is the dynamic polarizability of the Ry electron orbit. Indeed, despite experimental evidence~\cite{YouKnuAnd10}, the common wisdom has been that electrons in Rydberg states see trapping potentials that are essentially that of a free electron, and can only be trapped at laser intensity minima. In~\cite{TopDer13}, we demonstrated that this approximation is dramatically invalidated when the size of the Ry orbit is comparable to the lattice site extent.  There is a non-monotonic interplay between the lattice constant and the physical size of the Rydberg state. We explained this interplay with a simple toy model and showed that the size of the Rydberg orbit relative to the lattice  constant is what determines whether a Rydberg atom is trapped at the intensity minima or maxima.  The Rydberg electron trapping potential  is determined by the laser-intensity averaged ``landscaping" polarizability. Based on these ideas, Refs.~\cite{TopDer13,TopDer14} identified magic trapping 
conditions  for both alkali metal and alkaline earth atoms in infrared optical lattices. This paper may be considered as a natural extension of Ref.~\cite{TopDer14}, as here we require magic trapping of three levels.

As to the main results of this paper, we predict a triple magic wavelength for Yb atom where the $6s^2({^1}S_0)$ and $6s6p({^3}P_0)$ clock states experience the same optical potential as a $6sns({^1}S_0)$ Rydberg state. A recent realization of Yb lattice clock uses 759.35$\pm$0.02 nm optical lattice, which is only magic between these two clock states~\cite{BarHoyOat06}. We also surveyed other divalent atoms such as Ca, Mg, Hg and Sr but did not find any triple magic wavelengths in the far red detuned region. However, we do find that there are several triple magic wavelengths in the UV region. These wavelengths range from $\sim$220 nm to $\sim$400 nm, which can be generated using table-top setups for frequency-tripled Ti:sapphire, Nd:YAG and frequency-doubled dye lasers~\cite{Weber99}. 

A common source of gate errors is due to the finite lifetimes of Rydberg levels. This forces the gate operations to be performed on time scales much shorter than the Rydberg state  decay lifetimes~\cite{SafWal05}.  In addition to black-body-radiation-stimulated and spontaneous decays, optical trapping opens another loss channel: photoionization (PI) of Ry atom by the trapping laser. To assess the feasibility of the triple magic trapping  at both long (Yb) and short (Ca, Mg, Hg and Sr) wavelengths, we calculate the photoionization lifetimes using semiclassical expressions~\cite{DyaPank94,BetManYak12} and compare them to the natural lifetimes~\cite{Havey77,Jian96,FanXieZha01}. The PI lifetime calculations have been carried out for an effective intensity $I_{\rm{eff}} = \int I |\psi_{\rm{ho}}|^2 d^3\mathbf{r} = 10^4$ W/cm$^2$. This provides upper bounds for the PI lifetimes. We find that there are several Cooper minima in the wavelength ranges of interest and very long photoionization lifetimes (longer than the natural lifetimes) can be achieved at some of the triple magic wavelengths. For other decoherence mechanisms, we refer the reader to~\cite{KomTopKes16}, where fidelities are estimated for several sources of operational errors in an entangled network of Yb optical lattice clocks. 

The paper is organized as follows: In Sec~\ref{sec:form}, we describe the formalism. We then present results of numerical calculations of magic wavelengths, photoionization cross sections and photoionization lifetimes for Yb in Sec.~\ref{sec:yb_trip}. Finally in Sec.~\ref{sec:thz_trip}, we demonstrate that triple magic trapping condition can be met in the UV for Ca, Mg, Hg and Sr. We also give PI cross sections and PI lifetimes for these atoms in the wavelength ranges of interest. Finally, the conclusions are drawn in Sec.~\ref{sec:conc}. Atomic units are used throughout unless specified otherwise.

%--------------------------------------------------------------------------------------------
\section{Magic trapping divalent Rydberg atoms}\label{sec:form}
We start by briefly reviewing the ideas behind the magic trapping of Rydberg and ground state atoms in optical lattices. Details and derivations can be found in Ref.~\cite{TopDer13,TopDer14}. In the following, we assume a one-dimensional optical lattice  made up of two counter-propagating laser beams of frequency $\omega_{L}$ whose wave vector $k_{L}=2\pi/\lambda_{L}$ is along the $z$-axis. 

For a ground state atom, the trapping potential is $U(Z)=-\alpha_{g}(Z,\omega_{\rm L}) F^{2}_{\rm L}$ where $\alpha_{g}(Z,\omega_{\rm L})$ is the ground state dynamic polarizability and $F_{L}$ is the peak electric field strength of the lattice laser~\cite{Greiner03}. Typically, the size of the ground state atom is small compared to the size of the lattice constant, so the $Z$-dependence can be factored out and the dynamic polarizability can be written as 
\begin{equation}\label{eq:gs_polariz}
				\alpha_{g}(Z,\omega_{\rm L}) = \alpha_{g}(\omega_{\rm L}) \;\sin^2(k_{\rm L} Z) \;.
\end{equation}

In a Rydberg state, however, the size of the atom can be comparable or larger than the lattice constant $\lambda_{L}/2$, and the polarizability can be expressed as 
\begin{equation}\label{eq:ryd_polariz}
 \begin{split}
	 \alpha_{r}(Z, &\omega_{\rm L}) = 
			-\frac{1}{\omega_{\rm L}^2} 
				\langle nlm|\cos(2k_{\rm L} z_e)|nlm \rangle  \sin^2(k_{\rm L} Z) \\
			&+ \langle nlm|\sin^2(k_{\rm L} z_e)|nlm \rangle 
			+ \alpha_{\rm ion}(\omega_{\rm L}) \sin^2(k_{\rm L} Z) \;.
 \end{split}
\end{equation} 
Note that the first term depends on the position of the atom in the optical lattice the same way a ground state atom's polarizability does. The second term is independent of $Z$ and therefore does not play a role in trapping the atom in the optical lattice. The final term is the contribution from the residual ion seen by the Rydberg electron. We termed the $Z$-dependent contribution from the first term in~\eqref{eq:ryd_polariz} the landscaping polarizability as it modulates the free electron polarizability in accordance with the intensity landscape of the optical lattice~\cite{TopDer13}: 
\begin{equation}\label{eq:ry-lands}
	\alpha^{\rm lsc}_r(\omega_{\rm L}) \equiv -\frac{1}{\omega_{\rm L}^2} 
				\langle nlm|\cos(2k_{\rm L} z_e)|nlm \rangle \;. 
\end{equation}

In a $J=0$ divalent Rydberg state, where only one of the valence electrons is excited, $\alpha^{\rm lsc}_r(\omega_{\rm L})$ is added on top of the polarizability of the residual ion $\alpha_{\rm ion}(\omega_{\rm L})$ (e.g., Sr+), the polarizability $\alpha_{\rm core}(\omega_{\rm L})$ due to the contributions from core-excited states of doubly ionized atoms (e.g., Sr$^{2+}$), and a small term $\alpha_{\rm cv}(\omega_{\rm L})$ arising from excitations to occupied valence orbitals. 

The magic trapping condition is when the atoms experience the same optical trapping potential both in the ground and the Rydberg states, therefore when $\alpha_{g}(\omega_{L})=\alpha^{\rm lsc}_{r}(\omega_{L})$. In~\cite{TopDer13}, we found that this condition is satisfied for many laser wavelengths in the $1\lesssim \lambda_{L}\lesssim 10$ $\mu$m range for alkali Rydberg atoms with $n\lesssim 180$. In cold atom experiments which rely on trapping Rydberg atoms, working at such long wavelengths provides several compelling  advantages: in quantum gate experiments, the magic trapping scheme we describe renders turning off the trapping fields unnecessary whenever a gate operation is to be performed. Because the wavelengths are near the CO$_2$ laser band, the photon scattering and the ensuing motional heating is also reduced when compared to conventional traps near low lying resonances, alleviating an important source of decoherence. 

Same principles also apply to divalent atoms with two optically active electrons outside a closed shell, such as group-II atoms ({\it e.g.}, Mg, Ca, Sr) and group-II-like atoms such as Yb, Hg, Cd, and Zn. In~\cite{TopDer14}, we demonstrated that magic wavelengths can be found between the ground and singly-excited Rydberg states of Sr in the same wavelength range and developed a theoretical framework which takes into account the total angular symmetry of the wavefunction due to the existence of a spectator valence electron. 

Here we are interested in the divalent case where there is an additional level to the two clock states in the magic trapping condition. We consider a $\Lambda$-type system which consists of two low lying clock states and a single Rydberg state (Fig.~\ref{fig:diagram}(a)). The clock states used in optical lattice clocks are the ${^1}S_{0}$ and ${^3}P_{0}$ ground and excited states and we will consider ${^1}S_{0}$ Rydberg states although the same conditions apply equally well for the ${^1}P_{0}$ states as well~\cite{TopDer14}. 

The trapping potentials $U_{g}(Z)$, $U_{e}(Z)$ and $U_{r}(Z)$ (Fig.~\ref{fig:diagram}(b)) implicitly depend on time since the position $Z$ changes as the atom moves around inside the trap. To illustrate the idea behind magic trapping of a $\Lambda$-type system seen in Fig.~\ref{fig:diagram}(a), we will briefly ignore this implicit time-dependence and write the total wavefunction $|\psi\ket$ for the three level system as 
\begin{eqnarray}\label{eq:wf-1}
	|\psi\ket &=& |g\ket e^{-iU_{g}t} + |e\ket e^{-iU_{e}t} + |r\ket e^{-iU_{r}t} \\
						&=& e^{-iU_{g}t} \Big[ |g\ket + |e\ket e^{-i(U_{e}-U_{g})t} 
							+ |r\ket e^{-i(U_{r}-U_{g})t} \Big ] \nonumber \;,
\end{eqnarray}
where $|g\ket$, $|e\ket$, and $|r\ket$ refer to the ground (${^1}S_{0}$), excited (${^3}P_{0}$) clock states and the Rydberg state. The phases due to the trapping potential for each component are explicitly spelled out. 

Magic trapping between the ground $|g\ket$ and the excited $|e\ket$ states means that the atom experiences the same trapping potential in both states, therefore $U_{e}-U_{g}=0$. Therefore, in a magic wavelength optical lattice for the clock states, the total wave function~\eqref{eq:wf-1} can be written as  
\begin{equation}
	|\psi\ket = |g\ket + |e\ket + |r\ket e^{-i\Delta U_{r,g} t} \;,  \label{eq:wf-2}
\end{equation}
where we defined $\Delta U_{r,g}\equiv U_{r}-U_{g}$ and dropped the overall phase factor $\exp(-iU_{g}t)$ for brevity. The Ry atom will see the same trapping potential as the other two clock states when $\Delta U_{r,g}=0$. This helps eliminate dephasing due to the motion of the atom inside the optical lattice. We will refer this as the ``triple" magic trapping condition. Since $\Delta U_{r,g}= -\Delta\alpha(\omega_{\rm L}) F^{2}_{\rm L}$, it boils down to matching the clock state polarizabilities with the total polarizability of the Rydberg state,  
\begin{equation}
	\Delta\alpha_{g,e}(\omega_{L}) = \Delta \alpha_{g,r}(\omega_{L}) = 0 \;,  \label{eq:dalp_magic}
\end{equation}
where $\Delta\alpha_{j,k}=\alpha_{j}-\alpha_{k}$. The condition $\Delta\alpha_{g,e}(\omega_{L})=0$ is routinely exploited in optical lattice clocks and have become an integral part of the experimental tool box~\cite{DerKat11}. This condition in principle can be satisfied with arbitrary precision, {\it i.e.} $\Delta\alpha_{g,e}(\omega_{L})$ can be made negligibly small by varying $\omega_{\rm L}$. This is because monotonically varying $\alpha(\omega_{L})$ for the two clock states inevitably cross at some lattice wavelength. 

The situation is more complicated for the triple magic trapping because three curves may not cross at exactly the same point. Then one has to consider how small $\Delta\alpha_{g,r}$ can be made when $\Delta\alpha_{g,e}(\omega_{L})=0$. One also needs to establish a measure as to how small $\Delta\alpha_{g,r}$ needs to be before~\eqref{eq:dalp_magic} can be considered practically satisfied. We assess this condition by estimating the probability to transition out of the motional ground state due to the optical potential shift $\Delta U_{r,g}$ experienced by the Rydberg atom. In the following, we proceed with a derivation of an approximate expression for the transition probability. We also estimate the combined effects of $\Delta U_{r,g}$ and the photon recoil energy $E_{R}$ and compare it to the energy spacing between the motional states.

\subsection{Motional Excitation probability}\label{subsec:Pmot}
We begin by estimating the probability $P_{\rm mot}$ to transition from the ground state to an excited state of a harmonic oscillator (HO) potential as a result of a $\pi$-pulse driving the $|g\rangle \rightarrow |r\rangle$ transition. We will approximate the optical potential with a HO for the low-lying bound states. 

We describe the time-evolution of the system in the product basis of internal ($|c\rangle$ (clock, below we assume $|c\rangle=|g\rangle$ ) and $|r\rangle$ (Rydberg)), and motional $|\varphi_{n}\rangle$ states. The Hamiltonian of the system in the rotating wave approximation may be represented as 
\begin{equation}
H=\chi\left(  |g\rangle\langle r|+|r\rangle\langle g|\right)  +\frac{P^{2}%
}{2M}+U_{r}\left(Z\right)  |r\rangle\langle r|+U_{g}\left(Z\right)
|g\rangle\langle g|
\end{equation}
where $\chi\equiv\Omega/2$ and $\Omega$ is the Rabi frequency of the transition to the Ry state. The harmonic trapping potentials are 
\begin{align}
U_{r,g}\left(Z\right)   &  =\frac{1}{2}M\omega_{r,g}^{2}~Z^{2} \;, \\
\omega_{r,g}^{2}~  &  =\left(  \frac{2\pi}{\lambda_{L}}\right)  ^{2} \frac{2}{M}\alpha_{r,g}\left(  \omega_{L}\right)  F^2_{L} \;, 
\end{align}
where $F^2_{L}=I_{L}$ is intensity in a.u. and $M$ is the atomic mass.

We will treat the difference between the potentials $W\left(Z\right)=U_{r}\left(Z\right)-U_{g}\left(Z\right)$ as a perturbation: 
\begin{eqnarray}
&&W\left(Z\right) = \frac{1}{2}M~\Delta\omega_{\rm ho}^{2}~Z^{2} \;, \\
&&\begin{split}
	\Delta\omega_{\rm ho}^{2}  &  =\left[  \left(  \frac{2\pi}{\lambda_{L}}\right)^{2}\frac{2}{M} F^2_{L}\right]  
	\Delta\alpha_{r,g}   \;.
\end{split} 
\end{eqnarray}
where $\Delta\alpha_{r,g} = \alpha_{r}\left(\omega_{L}\right)  -\alpha_{g}\left(  \omega_{L}\right)$. With the ansatz 
\begin{equation}
\Psi\left(  r,Z,t\right)  =\Psi_{r}\left(Z,t\right)  |r\rangle+\Psi
_{g}\left(  Z,t\right)  |g\rangle \;,
\end{equation}
we obtain the following equations for the motional states $\Psi_{r,g}\left(Z,t\right)$ 
\begin{eqnarray*}\label{eq:pmot_schEq}
&	\begin{split}
			i\dot{\Psi}_{r}\left(Z,t\right) = &\chi\Psi_{g}\left(
				Z,t\right)  +\left\{  \tfrac{P^{2}}{2M}+U_{g}\left(Z\right)  \right\} \Psi_{r}\left(Z,t\right)  \\
				&+W\left(Z\right)  \Psi_{r}\left(Z,t\right) 
	\end{split}  \;, \\
&i\dot{\Psi}_{g}\left(Z,t\right) =\chi\Psi_{r}\left(
Z,t\right)  +\left\{  \tfrac{P^{2}}{2M}+U_{g}\left(Z\right)  \right\}
\Psi_{g}\left(Z,t\right)  \;.
\end{eqnarray*}

Next, we introduce the complete set of motional states for the ground state potential 
\begin{equation}
\left\{  \frac{P^{2}}{2M}+U_{g}\left(Z\right)  \right\}  |\Phi_{n}%
\rangle=E_{n}|\Phi_{n}\rangle \;. 
\end{equation}
For a harmonic potential, these are just the HO eigenstates. We expand $\Psi_{r}\left(Z,t\right)$ and $\Psi_{g}\left(Z,t\right)$ over  this complete set: 
\begin{align*}
\Psi_{r}\left(Z,t\right)   &  =\sum_{n}c_{n}^{r}\left(  t\right)
\exp\left(  -iE_{n}t\right)  |\Phi_{n}\rangle \;, \\
\Psi_{g}\left(  Z,t\right)   &  =\sum_{n}c_{n}^{g}\left(  t\right)
\exp\left(  -iE_{n}t\right)  |\Phi_{n}\rangle  \;.
\end{align*}
Substituting these into Eqs.~\eqref{eq:pmot_schEq} we arrive at 
\begin{align*}
i\dot{c}_{n}^{g}\left(  t\right)   &  =\chi c_{n}^{r}\left(  t\right) \;, \\
i\dot{c}_{n}^{r}\left(  t\right)   &  =\chi c_{n}^{g}\left(  t\right)
+\sum_{k}c_{k}^{r}\left(  t\right)  \exp\left(  -i\omega_{kn}t\right)  W_{nk} \;,
\end{align*}
where $W_{nk}=\langle \Phi_{n} |W| \Phi_{k} \rangle$. 

Since the system starts from the ground internal state and the ground motional eigenstate, we keep only $c_{0}^{g}\left(t\right)$. As a result of the $|g\rangle \rightarrow |r\rangle$ excitation, the system will be promoted to the desired $p=0$ motional state as well as to the undesired $p=2$ state.  We assume that the perturbation connects only to one state $|\Phi_{p}\rangle$ (for two harmonic potentials, the difference $W$ is also harmonic, therefore only the $p=2$ will be excited to the leading order due to the HO selection rules): 
\begin{align*}
i\dot{c}_{0}^{g}\left(  t\right)   &  =\chi c_{0}^{r}\left(  t\right) \;, \\
i\dot{c}_{0}^{r}\left(  t\right)   &  =\chi c_{0}^{g}\left(  t\right)
+c_{0}^{r}\left(  t\right)  W_{00}+c_{p}^{r}\left(  t\right)  \exp\left(
-i\omega_{p0}t\right)  W_{0p} \;, \\
i\dot{c}_{p}^{r}\left(  t\right)   &  =c_{0}^{r}\left(  t\right)  \exp\left(
-i\omega_{0p}t\right)  W_{p0}+c_{p}^{r}\left(  t\right)  W_{pp} \;. 
\end{align*}
We can absorb $W_{00}$ into the laser resonance frequency since this is just the energy shift of the transition, and use perturbation theory in $W$ 
\begin{align*}
i\dot{c}_{0}^{g}\left(  t\right)   &  =\chi c_{0}^{r}\left(  t\right) \;, \\
i\dot{c}_{0}^{r}\left(  t\right)   &  \approx\chi c_{0}^{g}\left(  t\right) \;, \\
i\dot{c}_{p}^{r}\left(  t\right)   &  \approx c_{0}^{r}\left(  t\right)
\exp\left(  -i\omega_{0p}t\right)  W_{p0}  \;. 
\end{align*}
Assuming that the system starts in the ground state, we obtain
\begin{align*}
c_{0}^{g}\left(  t\right)   &  =\cos\left(  \chi t\right) \;, \\
c_{0}^{r}\left(  t\right)   &  =\sin\left(  \chi t\right) \;, \\
i\dot{c}_{p}^{r}\left(  t\right)   &  \approx\sin\left(  \chi t\right)
\exp\left(  -i\omega_{0p}t\right)  W_{p0} \;,
\end{align*}
and
\begin{equation}
c_{p}^{r}\left(  \tau\right)  =\frac{1}{i}W_{p0}\int_{0}^{\tau}\sin\left(
\chi t\right)  \exp\left(  i\omega_{p0}t\right)  dt  \;.
\end{equation}
For a $\pi$ pulse, $\tau=\pi/\Omega,$ and
\begin{equation}
c_{p}^{r}\left(  \tau=\pi/\Omega\right)  =\frac{1}{i}W_{p0}\frac{\chi
+i\omega_{p0}e^{\frac{ii\omega_{p0}}{2\chi}}}{\chi^{2}-\omega_{p0}^{2}}  \;, 
\end{equation}
or the probability to end up in the excited motional state%
\begin{eqnarray}
P_{p}^{\rm mot} &=& \left\vert \frac{W_{p0}}%
{\omega_{p0}}\right\vert ^{2}G\left(  \chi/(\omega_{p0})\right) \;, \\
G\left(  \xi\right)  &=& \frac{1+\xi^{2}-2\xi\sin\left(  \frac{\pi}{2\xi}\right)
}{\left(  \xi^{2}-1\right)  ^{2}}  \;,
\end{eqnarray}
with $\xi=\left(  \Omega/2\right)  /\omega_{p0}$. For consistency, we have to
require that $P\ll1$, and this limit is attained for $\xi\gg1$ ($\Omega\gg
\Delta\omega_{\rm ho}$) and $G\left(  \xi\gg1\right) \approx 1/\xi^{2}$. This gives us  
\begin{eqnarray}
P_{p}^{\rm mot} &\approx&\left\vert \frac{W_{p0}}{\chi}\right\vert ^{2} \;,
\end{eqnarray}
which, for $p=2$, results in 
\begin{eqnarray}
P_{p=2}^{\rm mot} \equiv
	P_{\rm mot} \approx 
		\frac{1}{2}\left\vert \frac{\Delta\omega_{\rm ho}^{2}}{\Omega~\omega_{g}}\right\vert ^{2} \;. \label{eq:ppmot}
\end{eqnarray}

Clearly the probability goes down with the difference between the two harmonic potentials vanishing ({\it i.e.}, approaching the magic conditions) or since $\Delta\omega_{\rm ho}<$ $\omega_{g}$ by assumption that the two potentials are similar if $\Omega\gg\Delta\omega_{\rm ho}$ ({\it i.e.} the duration of the pulse is smaller than the oscillation period in the differential potential - as if the pulse were to take a perfect snapshot of the motional ground state). In terms of $\Delta\alpha_{r,g}$, Rabi frequency $\Omega$, lattice wavelength $\lambda_{L}$ and the intensity $F^2_{L}$, Eq.~\eqref{eq:ppmot} can be recast into the more useful form:  
\begin{equation}\label{eq:pmot}
	P_{\rm mot} \approx \frac{1}{2\Omega^2} \left(\frac{2\pi}{\lambda_{L}}\right)^{2} 
			\frac{2}{M} \frac{\Delta\alpha^2_{g,r}(\omega_{L})}{\alpha_{g}(\omega_{g})} 
			F^2_{L}  \;. 
\end{equation}

\subsection{Lamb-Dicke regime}\label{subsec:LD}
Experiments normally work in the Lamb-Dicke regime where the recoil energy is too small to induce transitions between the motional bound states: $E_{R}\ll \hbar\omega_{\rm ho}$. Here $E_{R}$ is the photon recoil energy and $\hbar\omega_{\rm ho}$ is the energy spacing between the motional states: 
\begin{eqnarray}\label{eq:omgHO}
	\omega_{\rm ho} &=& \frac{2\pi}{\lambda_{L}}\bigg( \frac{2 \alpha_g F_L^2 }{M} \bigg)^{1/2} \;, \\
	E_{\rm R} &=& \frac{k^2}{2M} = \frac{1}{2M} \bigg( \frac{2\pi}{\lambda_{\rm T}} \bigg)^2 \;. \label{eq:E-recoil}
\end{eqnarray}

The recoil can be from the clock transitions between the ${^1}S_{0}$ and ${^3}P_{0}$ states or the Ry transitions from the ground state. Here $\lambda_{L}$ and $\lambda_{\rm T}$ are the lattice laser and the transition wavelengths and $M$ is the mass of the atom. The transition wavelength $\lambda_{\rm T}$ refers to $\lambda_{\rm clock}$ when it is between the clock states $|g\ket$ and $|e\ket$, and to $\lambda_{\rm Ry}$ when between the ground and the Ry states (Fig.~\ref{fig:diagram}(a)). In obtaining~\eqref{eq:omgHO}, we approximated the optical potential with the harmonic oscillator for the low-lying bound states. An alternative way of looking at the condition $W\ll \Delta U_{r,g}$ is to compare the combined shift due to $E_{\rm R}$ and $\Delta U_{r,g}$ to $\hbar\omega_{\rm ho}$. We will do this estimate for the Yb atom below in the optical region of the spectrum. 

In the following sections, we show that the triple magic trapping condition can be approximately satisfied for the Yb atom in optical wavelengths and for various other divalent systems in the UV region. We estimate $P_{\rm mot}$ in each case to show that the discovered magic trapping conditions result in transition probabilities below $\sim$1\%.

\section{Triple magic trapping of Yb}\label{sec:yb_trip}
We start with a brief description of our calculations. The total dynamic polarizability of a $J=0$ divalent Ry atom  is~\cite{TopDer14} 
\begin{equation}\label{eq:div_alpha}
 \begin{split}
	\alpha^{J=0}_{n_{g}l_{g};n_{r}l_{r}}(\omega_{L}) = \alpha_{\rm ion}(\omega_{L}) 
			&+ \alpha^{\rm lsc,\; J=0}_{n_{r}l_{r}}(\omega_{L}) \\ 
			&+ \alpha_{\rm core}(\omega_{L}) + \alpha_{\rm cv}(\omega_{L}) \;.
 \end{split}
\end{equation}
Here $\alpha_{\rm core}(\omega_{L})$ is a contribution from the core-excited states of the doubly-ionized atoms (e.g. Yb$^{2+}$). This term is almost identical for both valence levels and is unimportant when considering the differential contributions to the polarizability. We also neglect $\alpha_{\rm cv}(\omega_{L})$ which arises from excitations to occupied valence orbitals as it is much smaller than $\alpha_{\rm core}(\omega_{L})$. The dynamic polarizability of the Yb$^{+}$ ion is given by 
\begin{equation}
\alpha_{\rm ion}(\omega) = \sum_{j} \frac{E_{6s}-E_{j}}{(E_{6s}-E_{j})^2 -\omega^2} 
	|\bra \psi_{\rm 6s} | \mathbf{D}|\psi_{j} \ket|^2 \;,
\end{equation}
where $\mathbf{D}$ is the electric dipole operator and $E_j$ are the ionic energy levels. We evaluate $\alpha_{\rm ion}(\omega)$ using a high-accuracy method~\cite{DerJohSaf99}. The result is in good agreement with the static polarizability from the literature~\cite{MitSafCla10} in the $\omega\rightarrow 0$ limit. 

The Rydberg landscaping polarizabilities $\alpha^{\rm lsc}_{r}(\omega_{L})$ are calculated using~\eqref{eq:ry-lands} where the Ry orbitals $|nlm\ket$ are obtained by integrating the time-independent Schr\"odinger equation. The ionic core potential for the Ry electron is modeled using the potential 
\begin{equation}\label{eq:model_pot}
	V(r) = -\frac{1}{r} -\frac{(Z_{a}-1)e^{-ar}}{r} + b e^{-cr} \;,
\end{equation}
where the constants $a$, $b$ and $c$ are determined by fitting the eigenenergies of $V(r)$ to the experimental energies for the $6sns({^1}S_{0})$ Ry series of Yb~\cite{MaeMatTak92,AymChaDel84} using a simulated annealing algorithm~\cite{GofFerRog94}. In the UV region, the dynamic polarizabilities are mainly the ion core polarizabilities. This is because the contribution from the Ry electron to the total polarizability is very small at short wavelengths and the details of the model potential does not matter~\cite{TopDer14}. In the long wavelength region, {\it e.g.} where the Yb triple magic wavelength lies, the dynamic polarizability oscillates and has nodes~\cite{TopDer13}. The positions of the nodes depend weakly on the details of the core, which is modeled by the small-$r$ part of the model potential~\eqref{eq:model_pot}. The small shifts in the positions of the nodes translate into small changes in the polarizabilities. The differences are qualitative and small enough not to change the fact that there is a triple magic wavelength around $\sim$550 nm to within a few nm. 

%----------------------------------------------------------
\begin{figure}[h!tb]
	\begin{center}
      \resizebox{\columnwidth}{!}{\includegraphics[angle=0]{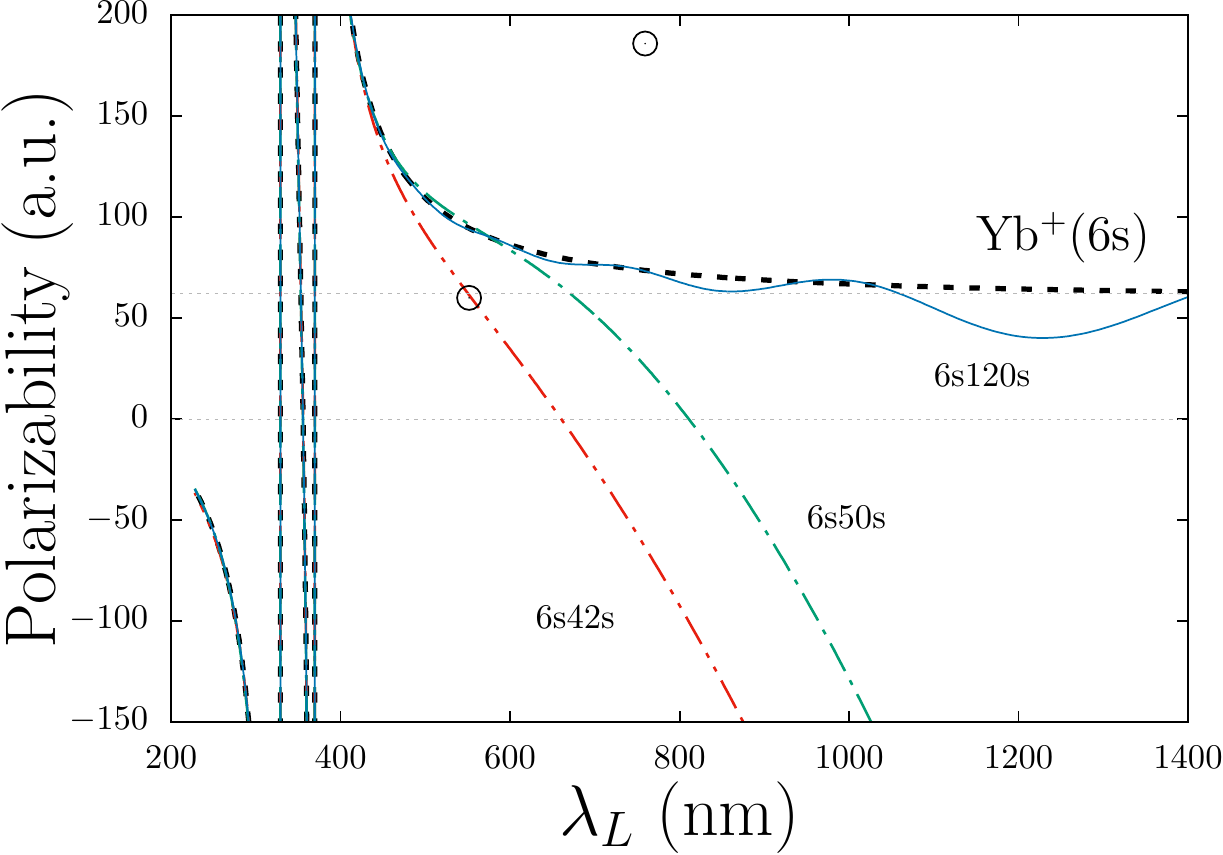}} 
	\end{center}
	\caption{(Color online) Yb triple magic trapping condition. The open circles represent the magic trapping conditions for the $6s^2({^1}S_0)$ and $6s6p({^3}P_0)$ clock states. The dashed black curve is the dynamic polarizability for the Yb$^{+}$ ion onto which the Rydberg landscaping polarizabilities are added to obtain the $6sns({^1}S_0)$ landscaping polarizabilities (solid blue, dot-dashed green, and dot-dot-dashed red curves). For $n=42$, the Rydberg state polarizability goes through one of the magic points at 551.5 nm for the ${^1}S_0$ and ${^3}P_0$ clock states, which constitutes a triple magic trapping condition. 
	}
	\label{fig:Yb_polariz}
\end{figure}
%----------------------------------------------------------
\begin{figure}[h!tb]
	\begin{center}
      \resizebox{\columnwidth}{!}{\includegraphics[angle=0]{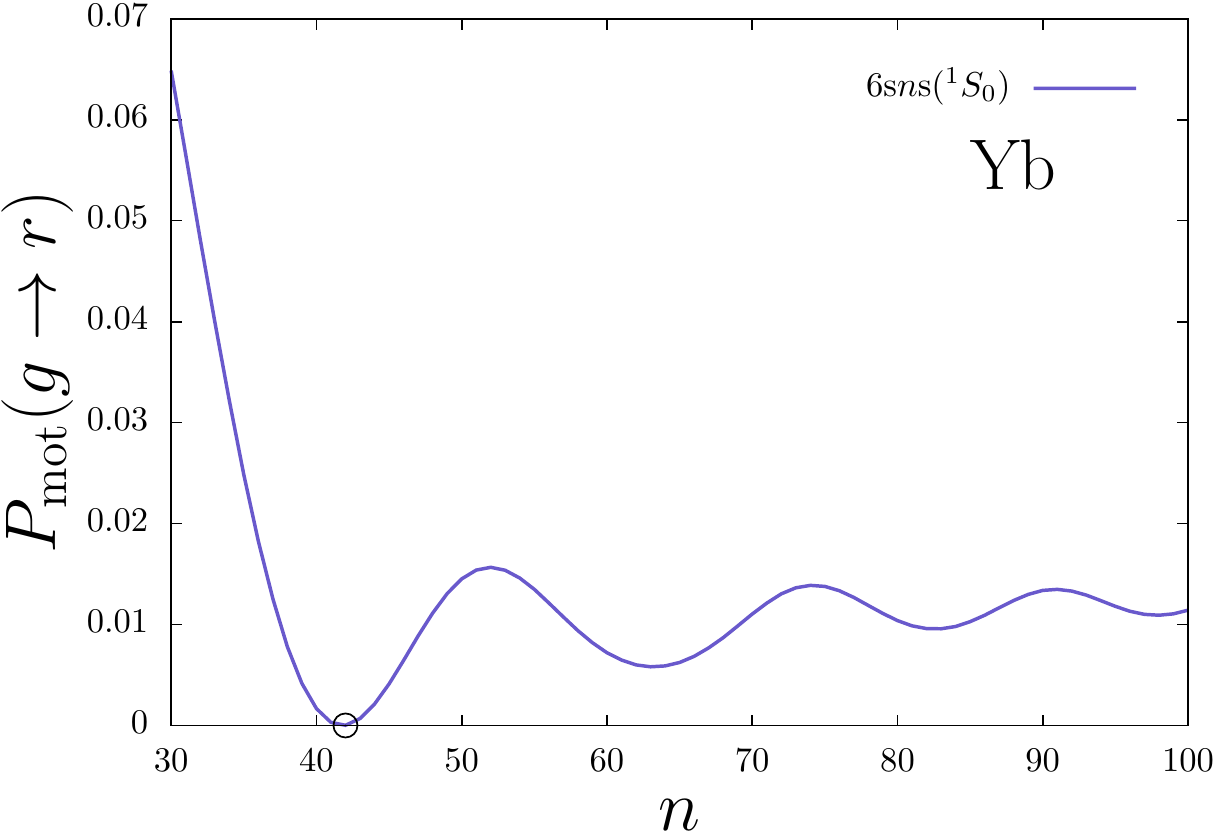}} 
	\end{center}
	\caption{(Color online) Excitation probability out of the motional ground state of the optical lattice potential for the 551.5 nm magic wavelength seen in Fig.~\ref{fig:Yb_polariz} for $\Omega=(2\pi)1$ MHz and lattice laser intensity of $10^{4}$ W/cm$^2$. The motional excitation probability is $\sim 10^{-5}$ when either of the clock states is excited to the $6s42s({^1}S_{0})$ Ry state. Probability approaches to $\sim$1\% in the limit $n\rightarrow \infty$. 
	}
	\label{fig:Yb_prob}
\end{figure}

As described in Sec.~\ref{sec:form}, we are looking for lattice wavelengths for which the polarizabilities of the $|g\ket=|6s^2({^1}S_{0})\ket$ and $|e\ket=|6s6p({^3}P_{0})\ket$ clock states match the polarizability of the $6sns({^1}S_{0})$ Ry state: 
\begin{equation}
	\alpha_{g}(\omega_{L}) = \alpha_{e}(\omega_{L}) \simeq \alpha^{({^1}S_{0})}_{6sns}(\omega_{L}) \;.  \label{eq:alp_magic2}
\end{equation}

Two cases for which $\Delta\alpha_{g,e}(\omega_{L})=0$ is satisfied are marked by empty circles in Fig.~\ref{fig:Yb_polariz}. At these magic points $\left( \lambda_{L},\; \alpha \right) = \left(551.5\;{\rm nm},\; 60\;{\rm a.u.}\right)$ and $\left(759.37\;{\rm nm},\; 186\;{\rm a.u.}\right)$~\cite{DzuDer10}, and the Yb atoms see the same trapping potential in both clock states. The $\alpha^{({^1}S_{0})}_{6sns}(\omega_{L})$ Ry state polarizabilities for $n=120$ (solid blue), 50 (dot-dashed green) and 42 (dot-dot-dashed red) are also plotted in Fig.~\ref{fig:Yb_polariz} along with $\alpha_{\rm ion}(\omega_{L})$ for the Yb$^{+}$ ion (dashed black curve). The condition $\Delta\alpha_{g,r}(\omega_{L})\simeq \Delta\alpha_{g,e}(\omega_{L})=0$ is best satisfied for the $n=42$ Ry state with $\Delta\alpha_{g,r}(\omega_{L})\simeq 1.5$ a.u.. 

\begin{figure}[h!tb]
	\begin{center}$
		\begin{array}{c}
    		\resizebox{\columnwidth}{!}{\includegraphics[angle=0]{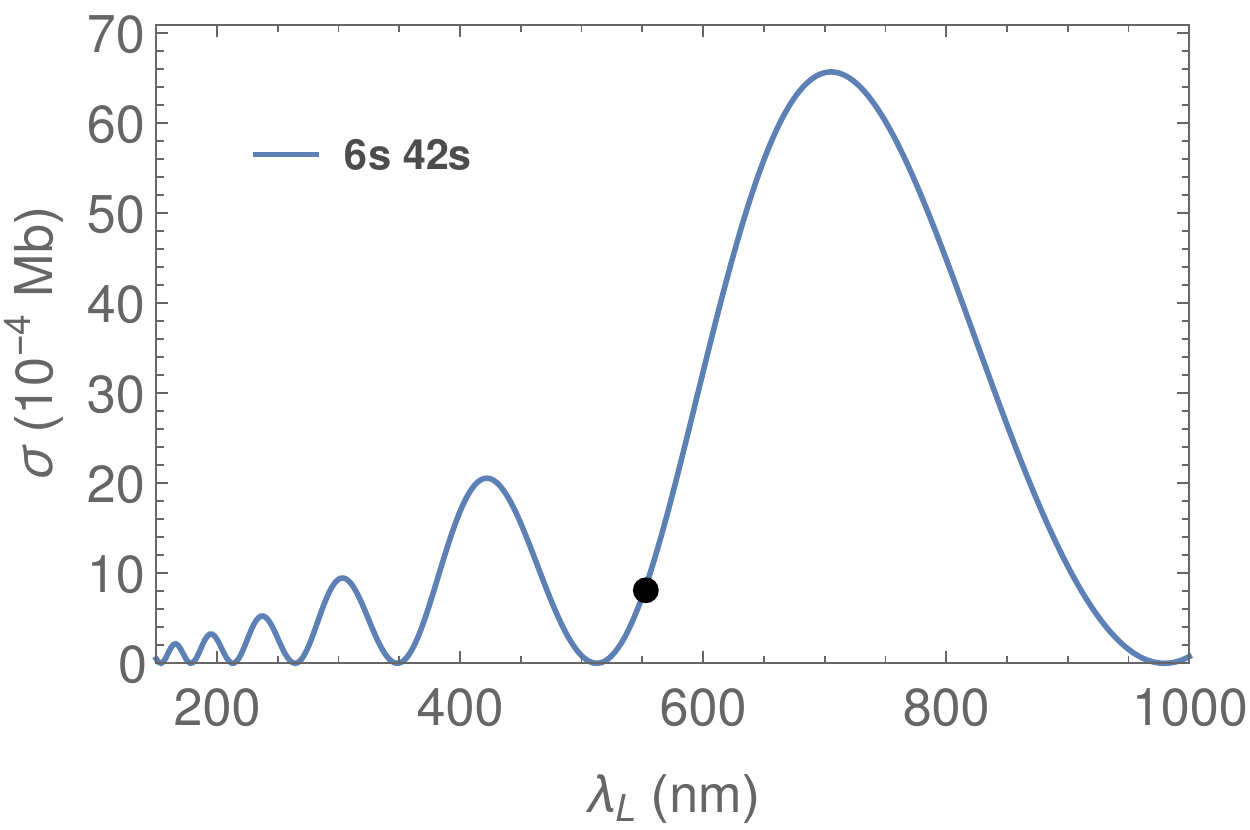}} \\ 
        \hspace{-0.1cm}
        \resizebox{0.98\columnwidth}{!}{\includegraphics[angle=0]{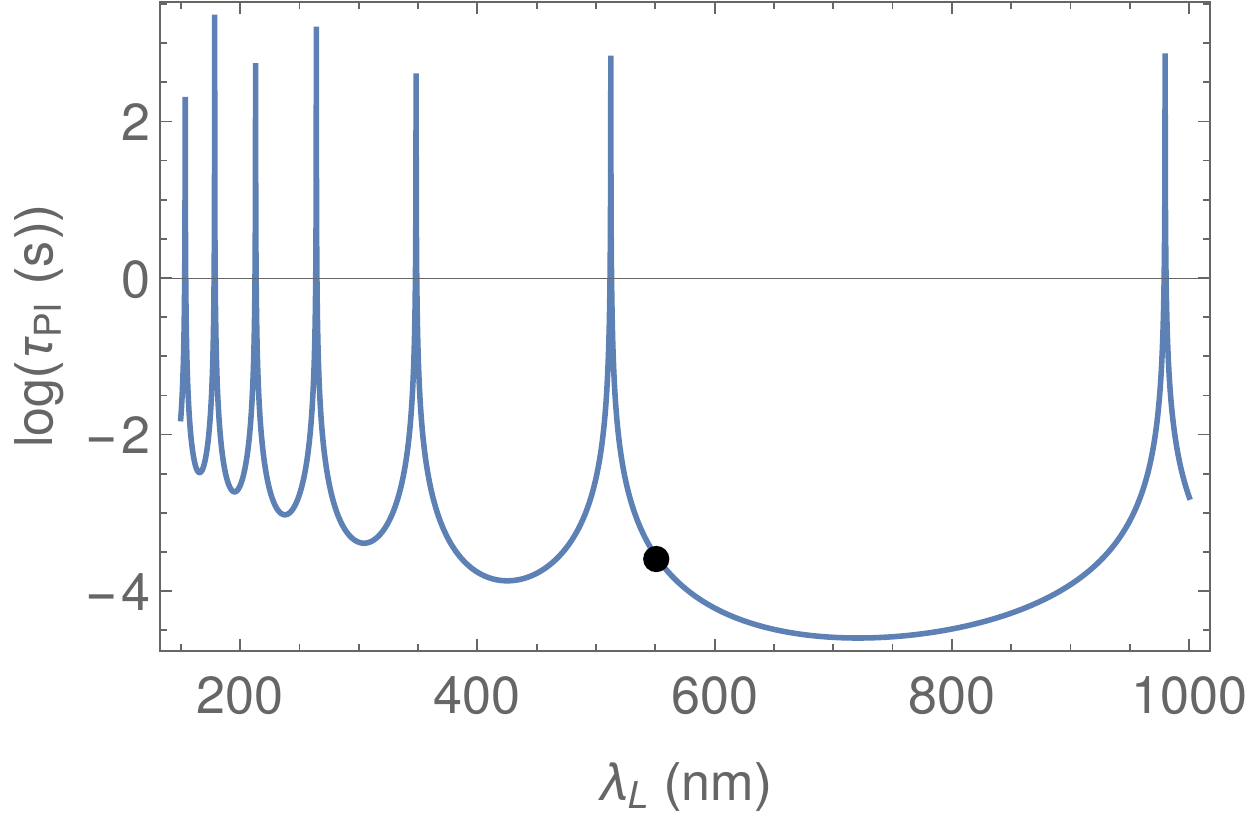}}
    \end{array}$
	\end{center}
	\caption{(Color online) (Upper panel) Photoionization cross section $\sigma$ for the $6s42s({^1}S_{0})$ Ry state of Yb. The cross section displays several Cooper minima below 1 $\mu$m. 
	(Lower panel) Logarithm of the photoionization lifetime of the $6s42s({^1}S_0)$ Rydberg state for which the triple magic trapping condition is satisfied when the trap intensity is $10^4$ W/cm$^2$. At the 551.5 nm triple magic wavelength, the photoionization lifetime is $\sim 2.67\times 10^{-4}$ s. 
	}
	\label{fig:Yb_CrossSec}
\end{figure}

The difference of $\Delta\alpha_{g,r}(\omega_{L})\simeq 1.5$ a.u. between the $n=42$ Ry state and the clock state polarizabilities translate into a probability of $P_{\rm mot}\sim 10^{-5}$ for excitation out of the motional ground state. For this Ry state, $P_{\rm mot}$ is small enough that the triple magic trapping condition is well suited for applications involving atomic clocks and QIP alike. The excitation probability estimated using Eq.~\eqref{eq:pmot} for various $6sns({^1}S_{0})$ Ry states of Yb using the 551.5 nm magic wavelength is plotted in Fig.~\ref{fig:Yb_prob}. It is clear that for all Ry states with $n\gtrsim 40$, the excitation probability is at the $\sim$1\% level, which maybe too large for atomic clocks but is reasonably small for QIP. 

The same $\Delta\alpha_{g,r}(\omega_{L})\simeq 1.5$ a.u. for the $n=42$ Ry state corresponds to $\Delta U_{r,g}/(\hbar\omega_{\rm ho})\approx 0.07$. At 551.5 nm lattice wavelength and 578 nm clock transition, $E_{R}/(\hbar\omega_{\rm ho})\approx 0.08$ and the Lamb-Dicke condition is well satisfied. This means that the total energy shift experienced by the $n=42$ Ry state is only 15\% of the motional energy level spacing compared to the 8\% change experienced by the clock states due to photon recoil alone. For reference, $E_{R}/(\hbar\omega_{\rm ho})\approx 0.054$ for Sr atomic clocks in a 814.427 nm magic wavelength optical lattice when intensity is 10$^4$ W/cm$^2$~\cite{TakHonHig05}. This corresponds to relative values of $E_{R}$ and $\hbar\omega_{\rm ho}$  such that the energy transfer due to photon recoil is only about 5\% of the motional energy level spacing. For our purposes, we can assume that $E_{R}$ can be tolerated to be a higher fraction of $\hbar\omega_{\rm ho}$ than 5\%, since quantum gates are 
not 
as susceptible to motional heating as the atomic clocks.

%--------------------------------------------------------------------------------------------
\begin{figure*}[h!tb]
		\begin{center}$
			\begin{array}{cccc}
         \resizebox{0.79\columnwidth}{!}{\includegraphics[angle=0]{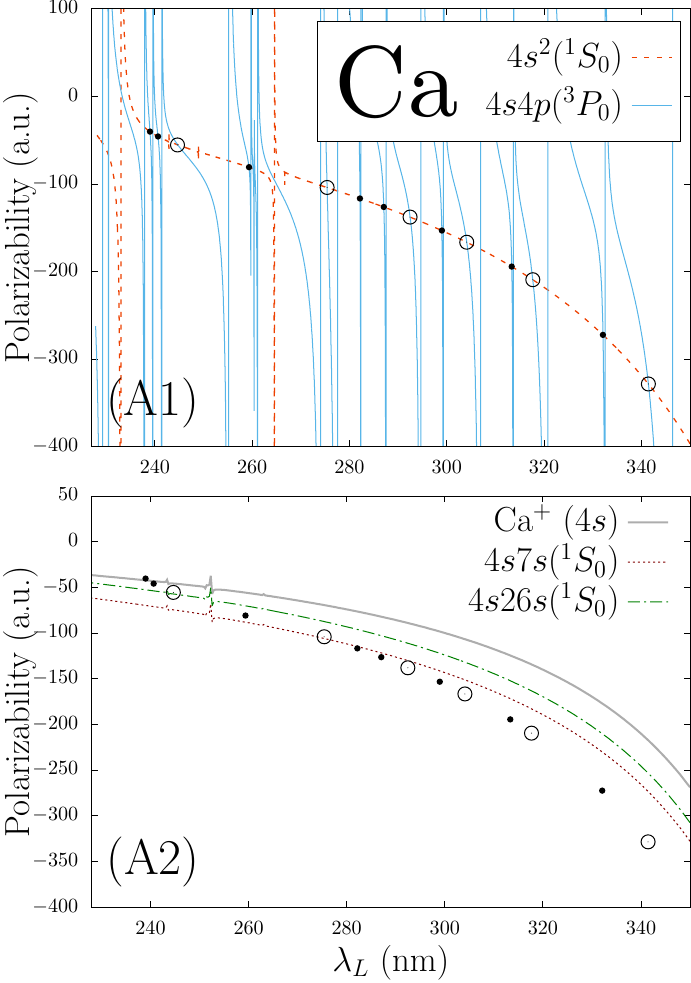}} & 
         \resizebox{0.81\columnwidth}{!}{\includegraphics[angle=0]{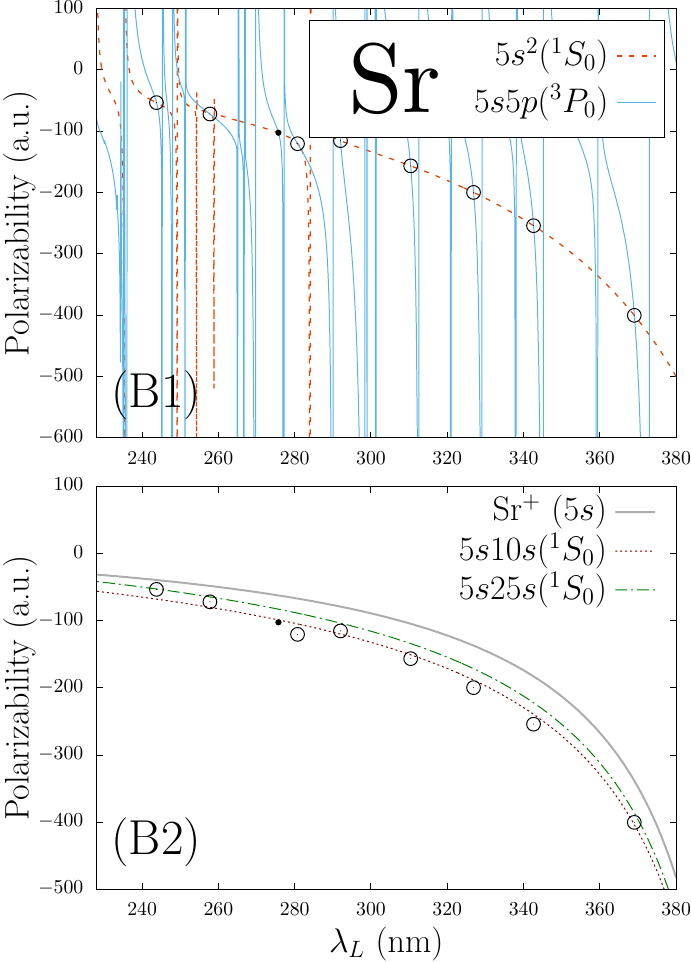}} \\ 
         \resizebox{0.82\columnwidth}{!}{\includegraphics[angle=0]{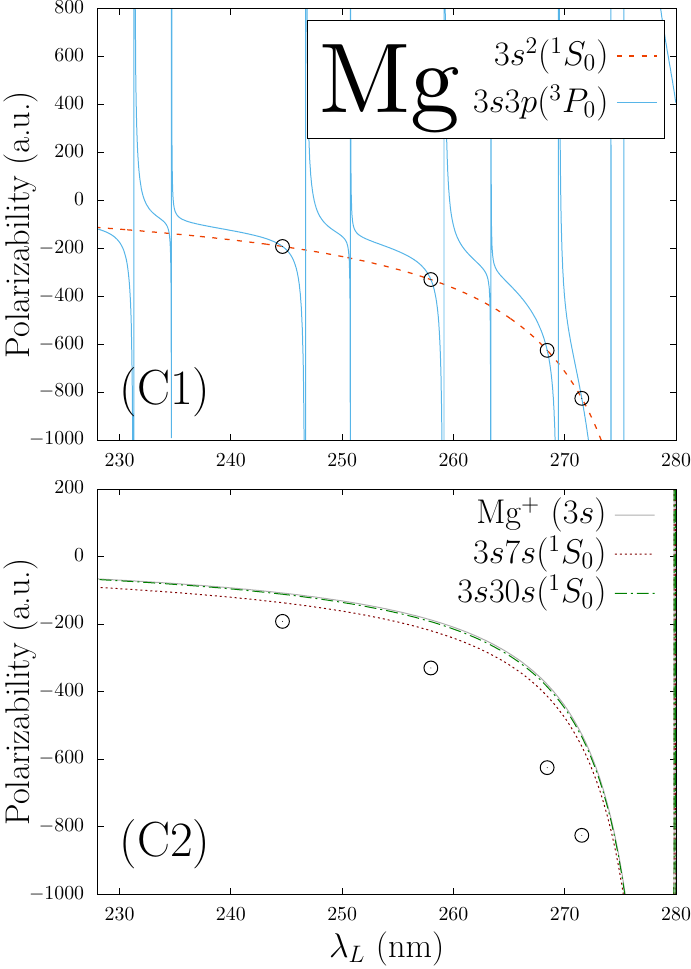}} & 
				 \resizebox{0.82\columnwidth}{!}{\includegraphics[angle=0]{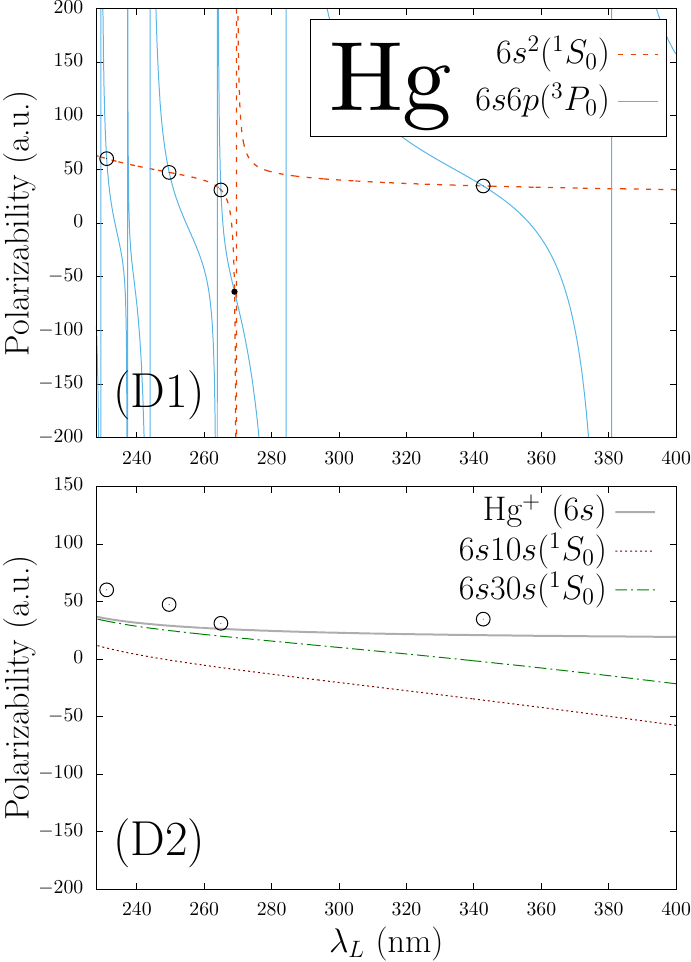}} 
  	  \end{array}$
		\end{center}
	\caption{(Color online) (Panels A1, B1, C1 and D1) Magic wavelengths for trapping Ca, Mg, Hg and Sr atoms in clock states in the UV region. The dashed red curves are $\alpha_{g}(\omega_{L})$ for the ${^1}S_{0}$ lower clock state and the solid blue curves are $\alpha_{e}(\omega_{L})$ for the ${^3}P_{0}$ upper clock state. The off-resonant magic trapping conditions, where $\Delta\alpha_{g,e}(\omega_{L})=0$, are labeled with open circles. Other points where $\Delta\alpha_{g,e}(\omega_{L})=0$ are also labeled with solid black points although these are too close to resonances for us to consider magic wavelengths. 
	(Panels A2, B2, C2 and D2) Open circles from the upper panels with $\alpha_{\rm ion}(\omega_{L})$ (solid gray) and $\alpha^{({^1}S_{0})}_{nsn's}(\omega)$ for two Ry states (dashed red and dot-dashed green). All points for which $\Delta\alpha_{g,e}(\omega_{L})=0$ lie below $\alpha_{\rm ion}(\omega_{L})$ and there are several wavelengths where the triple magic trapping condition can be satisfied. 
	}
	\label{fig:div_polariz}
\end{figure*}
%--------------------------------------------------------------------------------------------

The existence of the triple magic trapping condition between the $6s^2({^1}S_{0})$, $6s6p({^3}P_{0})$ clock states and the $6s42s({^1}S_{0})$ in Fig.~\ref{fig:Yb_polariz} relies on the existence of a magic trapping condition for the two clock states alone, whose polarizability lies below that of the Yb$^{+}$ ion. In fact, {\it a triple magic trapping condition, such as the one in Fig.~\ref{fig:Yb_polariz}, can always be found if $\alpha_{g}(\omega^{\text{*}}_{L})$ is below $\alpha_{\rm ion}(\omega_{L})$ at the magic wavelength $\lambda^{\text{*}}_{L}=2\pi/\omega^{\text{*}}_{L}$ for which $\alpha_{g}(\omega^{\text{*}}_{L})=\alpha_{e}(\omega^{\text{*}}_{L})$. } Unfortunately, among the divalent atoms we surveyed (Ca, Mg, Hg and Sr), we find that Yb is the only one for which this condition can be satisfied in the optical wavelengths. 

We also estimated the photoionization cross sections and lifetimes in the $6s42s({^1}S_{0})$ Rydberg state for lattice wavelengths up to 1 $\mu$m using analytical expressions obtained through semiclassical approximations. We refer the reader to Refs.~\cite{DyaPank94} for a detailed derivation and to~\cite{BetManYak12} for a description of its implementation in terms of closed form analytical functions. Since the photoionization cross sections involve bound-continuum matrix elements, the quantum defects need to be extrapolated into the continuum, especially in the wavelengths regions involving Cooper minima~\cite{Seaton57}. The cross sections and the corresponding PI lifetimes are shown in Fig.~\ref{fig:Yb_CrossSec}. There are several Cooper minima in the region $\lambda_{L}<1$ $\mu$m, which correspond to the long-lived peaks in the lower panel. At the 551.5 nm triple magic wavelength seen in Fig.~\ref{fig:Yb_polariz}, the photoionization lifetime is $\sim$267 $\mu$s. The radiative lifetime for the same state 
is $\sim$46 $\mu$s at 300 K~\cite{FanXieZha01}. This suggests that the lifetime of the $6s42s({^1}S_{0})$ Ry state trapped in a triple magic optical lattice is limited by its radiative lifetime in room temperature. 

%--------------------------------------------------------------------------------------------
\section{Triple magic trapping in the UV: Ca, Mg, Hg and Sr}\label{sec:thz_trip}
Although we were able to find triple magic wavelengths for Yb in the optical part of the spectrum, there are several cases for other divalent atoms for which the condition $\Delta\alpha_{g,r}\approx 0$ can only be satisfied in the UV region. In this section, we survey Ca, Sr, Mg and Hg for wavelengths below 400 nm and discuss some wavelengths where the condition $\Delta\alpha_{g,r}\approx 0$ is approximately satisfied. We estimate excitation probabilities $P_{\rm mot}$ and give semiclassical PI cross sections and lifetimes for atom and compare to natural lifetimes when experimental data is available. 

The results we discuss below are based on data from the polarizability plots seen in Fig.~\ref{fig:div_polariz}. The upper panel (A1, B1, C1, and D1) for each atom in Fig.~\ref{fig:div_polariz} shows the dynamic polarizabilities for the lower $|g\ket=|ns^2({^1}S_{0})\ket$ and upper $|e\ket=|nsn'p({^3}P_{0})\ket$ clock states in the $\lambda_{L}\lesssim400$ nm region. Open circles represent the best candidates for the triple magic trapping conditions where the dynamic polarizabilities for the clock states match, i.e. $\Delta\alpha_{g,e}(\omega_{L})=0$. We label other points where this condition is also satisfied (small black points), however, these points are too close to resonances for us to consider useful for trapping. The lower panels (A2, B2, C2, and D2) plot these same points with the ionic polarizabilities $\alpha_{\rm ion}(\omega_{L})$ (solid gray) and two Ry state polarizabilities for which the triple magic trapping can be best attained (red dotted and green dash-dotted lines). Below we discuss 
how well the triple 
magic trapping condition $\Delta\alpha_{g,r}(\omega_{L})\approx 0$ is satisfied for each atom in Fig.~\ref{fig:div_polariz}. 

\subsection{Calcium}\label{subsec:ca}
Most of the magic wavelengths for which $\Delta\alpha_{g,e}=0$ from panel (A1) can be treated as triple magic wavelengths to a good degree. This is facilitated by the fact that $\alpha_{\rm ion}(\omega_{L})$ lies above all the magic wavelengths from the panel (A1). The smaller the $n$ the closer the total Ry state polarizability to $\alpha_{\rm ion}(\omega_{L})$. As $n$ is decreased, Ry state polarizability becomes more negative and sweeps out a region whose size is comparable to the spread of the magic points from the upper panel (empty circles). The shortest two magic wavelengths seem particularly promising as $P_{\rm mot}\simeq 7\times 10^{-4}$ and $3.6\times 10^{-4}$ can be achieved at these points for the $4s26s({^1}S_{0})$ and $4s7s({^1}S_{0})$ states, respectively. 

The probabilities for transitioning out of the motional ground state ($P_{\rm mot}$) are all below $\sim$10\% for all the magic wavelengths (empty points) in panel (A2) for the $4s26s({^1}S_{0})$ Ry state. For this state, the largest one is $P_{\rm mot}\simeq 0.13$ at the longest magic wavelength at 341.4 nm. The estimates of $P_{\rm mot}$ for the $4s7s({^1}S_{0})$ state are less than 0.03 at most magic wavelengths with the exception of the shortest and the longest ones at which $P_{\rm mot}\simeq 0.09$ and 0.07, respectively. 

Long PI lifetimes can be achieved in this wavelength range due to the small PI cross sections (see Fig.~\ref{fig:CrossSec}). Although there are no Cooper minima for Ca in the $\lambda_{L}$ range we consider in Fig.~\ref{fig:div_polariz}, Fig.~\ref{fig:lifetimes} suggests that PI lifetimes above 100 $\mu$s can be achieved, which is much longer than the natural lifetimes of the $4s7s({^1}S_{0})$ ($\sim$62 ns) and $4s26s({^1}S_{0})$ ($\sim$8 $\mu$s) states~\cite{Havey77}.

\subsection{Strontium}\label{subsec:sr}
Sr is the most promising candidate for triple magic trapping in all the atoms we surveyed in the UV wavelength region. All eight wavelengths labeled with open circles in the upper panel for Sr in Fig.~\ref{fig:div_polariz} can potentially be considered a triple magic wavelength depending on $n$ of the Ry state $5sns({^1}S_{0})$. Most of the $\Delta\alpha_{g,e}(\omega_{L})=0$ magic wavelengths from (B1) can be made to be triple magic by an appropriate choice of $n$. For all the magic wavelengths in panel (B2), $P_{\rm mot}\lesssim 0.01$ for both $5s10s({^1}S_{0})$ and $5s25s({^1}S_{0})$ Ry states of Sr suggesting motional excitation probabilities below 1\%. The lowest $P_{\rm mot}$ are $3.6\times 10^{-5}$ and $9\times 10^{-5}$ for the $5s10s({^1}S_{0})$ and $5s25s({^1}S_{0})$ states at the shortest and the longest wavelengths, respectively. 

\begin{figure}[h!tb]
	\begin{center}
		\resizebox{\columnwidth}{!}{\includegraphics[angle=0]{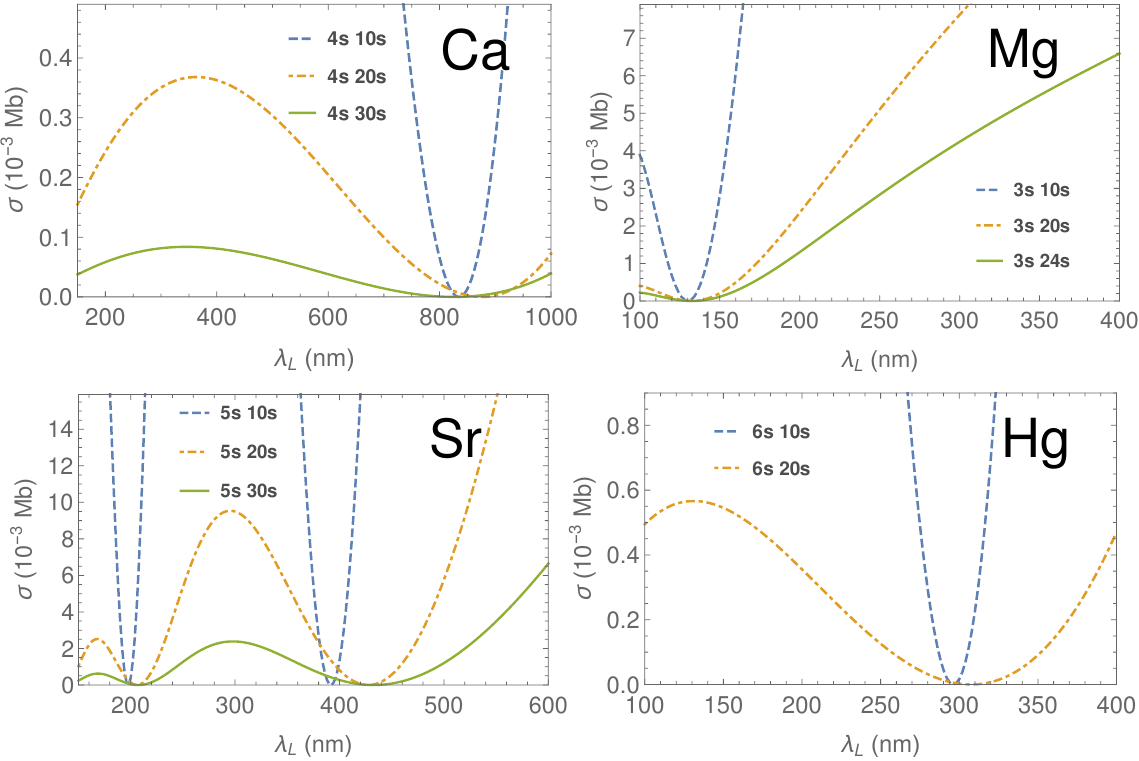}}
	\end{center}
	\caption{(Color online) Photoionization cross sections for Ca, Mg, Hg and Sr out of $nsn's({^1}S_{0})$ Ry states for few $n'$. There are several Cooper minima in the $\lambda_{L}<1$ $\mu$m wavelength range which can help minimize ionization due to lattice lasers for optical trapping of Ry atoms in the UV region. 
	}
	\label{fig:CrossSec}
\end{figure}

\begin{figure}[h!tb]
	\begin{center}
		\resizebox{\columnwidth}{!}{\includegraphics[angle=0]{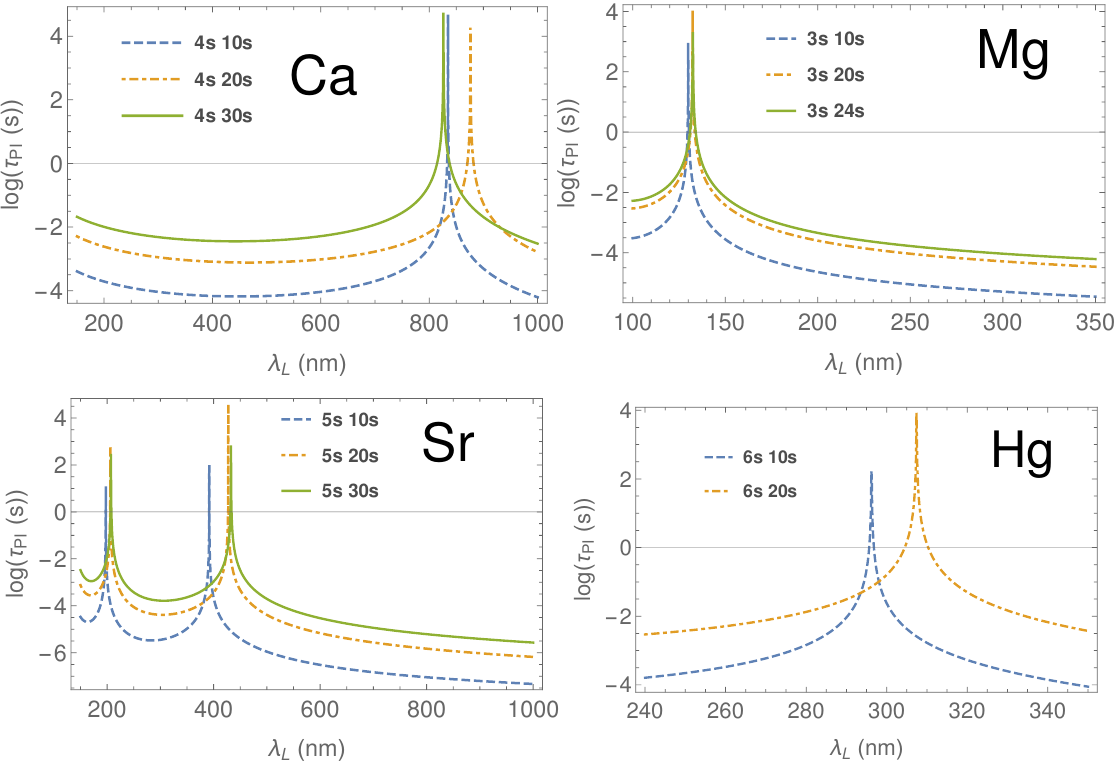}}
	\end{center}
	\caption{(Color online) Logarithm of the photoionization lifetimes corresponding to the cross sections seen in Fig.~\ref{fig:CrossSec}. Long-lived peaks result from the Cooper minima. The intensity of the lattice lasers is taken to be $10^{4}$ W/cm$^2$ which corresponds to a 10 $\mu$K deep trap for a 813 nm magic wavelength lattice for Sr optical lattice clock~\cite{TakHonHig05}. 
	}
	\label{fig:lifetimes}
\end{figure}

Due to the existence of Cooper minima around $\lambda_{L}\approx 200$ nm and 400 nm, long lived Ry states may be chosen by appropriating $n$ for $\lambda_{L}<500$ nm. For example, Sr panel of Fig.~\ref{fig:CrossSec} suggests that there is $\sim$50 nm shift in the position of the Cooper minima around $\lambda_{L}\approx 400$ nm between $n=10$ and $n=30$. Nevertheless, in the wavelength range of Fig.~\ref{fig:div_polariz}, PI lifetimes are longer than 100 $\mu$s for 10$^4$ W/cm$^2$ lattice laser intensity while the natural lifetimes for the $n=10$ and $n=25$ states are $\sim$300 ns and $\sim$10 $\mu$s respectively~\cite{Jian96}. 

%--------------------------------------------------------------------------------------------

%--------------------------------------------------------------------------------------------

\subsection{Magnesium}\label{subsec:mg}
Unfortunately, Mg is not as promising a candidate as Ca or Sr for triple magic trapping. For the four $\Delta\alpha_{g,e}=0$ magic wavelengths labeled in panel (C1), the smallest motional excitation probability $P_{\rm mot}$ is 38\% for the $3s7s({^1}S_{0})$ Ry state whereas it is 80\% for the $3s30s({^1}S_{0})$ state which occurs at the lowest magic wavelength in panels (C1) and (C2). At longer wavelengths shown, $P_{\rm mot}$ is larger for both Ry states. The PI lifetimes do not offer much comfort either as the PI cross sections are large in this wavelength region (Fig.~\ref{fig:CrossSec}). The PI lifetimes vary between 1 and 100 $\mu$s whereas the natural lifetime for the $3s30s({^1}S_{0})$ state is $\sim$50  $\mu$s~\cite{Jian96}. 

\subsection{Mercury}\label{subsec:hg}
Mercury is somewhat more amenable than Mg. The smallest $P_{\rm mot}$ is 3\% for the $6s30s({^1}S_{0})$ state at 249.6 nm and the largest $P_{\rm mot}$ is 21\% for the $6s10s({^1}S_{0})$ state at 342.8 nm. At the magic wavelengths shown in panels (D1) and (D2),  $P_{\rm mot}$ is mostly around 10\% level for both of these Ry states. The photoionization lifetimes are all longer then 100 $\mu$s in this wavelength range (Fig.~\ref{fig:lifetimes}) due to a Cooper minimum around 300 nm (Fig.~\ref{fig:CrossSec}).

\section{Conclusions}\label{sec:conc}
We have surveyed a set of divalent atoms to assess the possibility of triple magic trapping cold Rydberg atoms in optical lattices. This condition allows for trapping $\Lambda$-type atomic systems such that all the differential shifts due to the atomic motion inside the optical lattice vanish between ${^1}S_{0}$ and ${^3}P_{0}$ clock states together with that for a ${^1}S_{0}$ Rydberg state. For such a condition to be realized, there has to be magic wavelengths for the two clock states alone for which dynamic polarizability at the magic wavelength is below that of the ionic ground state. Since the ion polarizability and the Ry landscaping polarizabilities simply add for a $J=0$ Ry state, this guarantees at least one Ry state with approximately the same polarizability as the two clock states at this wavelength. 

We inspected Ca, Mg, Sr, Hg and Yb atoms both in the optical wavelengths as well as in the UV region. To assess the quality of the triple magic trapping conditions in each case, we estimated the excitation probabilities out of the motional ground state of the optical trap resulting from the $|g\rangle \rightarrow |r\rangle$ excitation. In the optical wavelengths, we only found one viable case for Yb where the triple magic trapping condition can be efficiently satisfied for one Ry state. We also estimated the PI cross section and lifetime of this state and concluded that it is much longer than its natural lifetime. 

We found that Sr and Ca are reasonably good candidates in the UV region. There are several magic wavelength with $\Delta\alpha_{g,e}(\omega_{L})=0$ for $\lambda_{L}\lesssim 400$ nm, where the probability for transitioning out of the motional ground state is reasonably small, at 1\% level. The PI lifetimes for Sr and Ca are also favorable in this wavelength region as they are longer than 100 $\mu$s. This is substantially longer than the natural lifetimes of the Ry states we consider.

\section{Acknowledgements}
This work was supported by the National Science Foundation (NSF) Grant No. PHY-1212482. A.D. was also supported by the Simons foundation as a Simons fellow in theoretical physics. T.T. and A.D. would like to thank the Institute for Theoretical Atomic, Molecular and Optical Physics (ITAMP) and the Harvard University Physics Department for their hospitality, where part of this work was carried out.

\bibliographystyle{prl2012}
\bibliography{TripleMagic,TripleMagic2}

\end{document}